\documentclass[a4paper]{jpconf}

\usepackage{graphicx}

\newcommand{\be}{\begin{equation}}
\newcommand{\ee}{\end{equation}}

\newcommand{\dlt}{\delta}
\newcommand{\bt}{\beta}
\newcommand{\vp}{\varphi}
\newcommand{\ep}{\varepsilon}
\newcommand{\al}{\alpha}
\newcommand{\ra}{\rightarrow}
\newcommand{\sgm}{\sigma}

\newcommand{\Gm}{\Gamma}

\newcommand{\lbd}{\lambda}

\newcommand{\cH}{{\cal H}}

\newcommand{\rgl}{\rangle}
\newcommand{\lgl}{\langle}

\begin{document}
\title{Quantum probabilities and entanglement for multimode
quantum systems}

\author{V I Yukalov$^{1,2}$, E P Yukalova$^{1,3}$ 
and D Sornette$^{1,4}$}

\address{
$^1$DMTEC, ETH Z\"urich, Swiss Federal Institute of Technology, 
Z\"urich CH-8092, Switzerland} 

\address{
$^2$Bogolubov Laboratory of Theoretical Physics, 
Joint Institute for Nuclear Research, Dubna 141980, Russia} 

\address{
$^3$Laboratory of Information Technologies, 
Joint Institute for Nuclear Research, Dubna 141980, Russia}

\address{
$^4$Swiss Finance Institute, c/o University of Geneva,
40 blvd. Du Pont d'Arve, CH 1211 Geneva 4, Switzerland}

\ead{yukalov@theor.jinr.ru}

\begin{abstract}
Quantum probabilities are defined for several important physical cases
characterizing measurements with multimode quantum systems. These are the
probabilities for operationally testable measurements, for operationally
uncertain measurements, and for entangled composite events. The role of
the prospect and state entanglement is emphasized. Numerical modeling
is presented for a two-mode Bose-condensed system of trapped atoms. The
interference factor is calculated by invoking the channel-state duality.
\end{abstract}

\section{Introduction}

Multimode quantum systems are of great interest because of their importance
for a variety of applications, such as quantum electronics and quantum
information processing [1]. There exists a number of multimode-system types
of different physical nature. These, for instance, can be atomic systems
with several excited electron levels, molecular systems with several
roto-vibrational modes, quantum dots with several populated exciton levels,
spin assemblies with several spin projections, trapped ions and trapped
neutral atoms with several energy states, and so on [2].

For the purpose of quantum information processing and quantum computing, it
is necessary to have well defined notions of quantum probabilities, including
the probability of composite events. Quantum probabilities of separate events,
describing quantum measurements, have been defined by von Neumann [3]. The
development of this notion can be found in the recent reviews [4,5]. Note that,
starting from von Neumann [3], the theory of quantum measurements is considered
as an analogue of the quantum decision theory [6,7]. Bohr [8-10] has argued
that quantum theory is an appropriate tool for describing human decision making.
It has been shown [11-15] that quantum decision theory, for both quantum
measurements as well as for human decision making, can be developed on the same
grounds employing quantum techniques. Mathematical problems appearing when introducing
quantum probabilities for composite events are discussed in Ref. [16].

In the present paper, after recalling the notion of the quantum probabilities
for operationally testable measurements (Sec. 2), we introduce the notion of
operationally uncertain measurements (Sec. 3). Then, in Sec. 4, following
the theory of Ref. [16], we define the joint quantum probabilities for composite
events, emphasizing the application to measurements under uncertainty. In Sec. 5,
we discuss how one can estimate the average value of the interference factor.
We show that this value, under a rather general assumption of its symmetric
distribution, equals $1/4$. The necessity of entanglement for the appearance
of coherent interference is stressed in Sec. 6. An example of calculating the
interference factor, employing the channel-state duality, for the case of a
trapped Bose-Einstein condensate with generated coherent modes, is given in
Sec. 7, also containing a brief conclusion.

\section{Operationally testable measurements}

First, let us recall the definition of quantum probabilities for separate
measurements representing quantum events [3]. Quantum events, obeying the
Birkhoff-von Neumann quantum logic [17] form a non-commutative
non-distributive ring $\mathcal{R}$. The nonempty collection of all subsets
of the event ring $\mathcal{R}$, including $\mathcal{R}$, which is closed
with respect to countable unions and complements, is the event sigma algebra
$\Sigma$. The algebra of quantum events is the pair $\{\Sigma, \mathcal{R}\}$
of the sigma algebra $\Sigma$ over the event ring $\mathcal{R}$.

Observables in quantum theory are represented by self-adjoint operators
$\hat{A}$ on a Hilbert space $\mathcal{H}$, pertaining to the algebra of
local observables $\mathcal{A}$. From the eigenproblem
\be
\label{1}
 \hat A | n \rgl = A_n | n \rgl \;  ,
\ee
one defines the projectors
\be
\label{2}
\hat P_n \equiv   | n \rgl \lgl n | \;  ,
\ee
allowing for the spectral decomposition
\be
\label{3}
 \hat A = \sum_n A_n \hat P_n \; .
\ee
The projectors are summed to the unity operator in $\mathcal{H}$:
\be
\label{4}
 \sum_n \hat P_n  = \hat 1_\cH \;   .
\ee

The operator spectrum $\{ A_n \}$ can be discrete or continuous, degenerate
or not. In what follows, we assume, for simplicity, a discrete non-degenerate
spectrum. The generalization to an arbitrary spectrum is straightforward.
Assuming a non-degenerate spectrum, we keep in mind the von Neumann
suggestion [3] of avoiding degeneracy by lifting it with additional small
terms breaking the operator symmetry responsible for the spectrum degeneracy,
and at the end removing these terms. In contrast, a discrete spectrum is typical
of finite quantum systems [2].

A physical system is characterized by a statistical operator $\hat{\varrho}$,
called the {\it system state}, which is a trace-class positive operator
normalized to one, so that the trace over $\mathcal{H}$ is Tr$\hat{\varrho} = 1$.
As a result of a measurement with an operator $\hat{A}$, one can observe an
eigenvalue $A_n$, which can be termed the event $A_n$. The probability of this
event is
\be
\label{5}
 p(A_n) \equiv {\rm Tr}\hat\rho \hat P_n \equiv
\lgl \hat P_n \rgl \; ,
\ee
with the trace over $\mathcal{H}$. By this definition,
$$
 \sum_n p(A_n) = 1 \; , \qquad 0 \leq p(A_n) \leq 1 \;  ,
$$
hence the family $\{ p(A_n) \}$ forms a probability measure. According to the
Gleason theorem [18], this measure is unique for a Hilbert space of
dimensionality larger than two.

In the theory of quantum measurements, the projectors $\hat{P}_n$ play the role
of observables, so that, for an event $A_n$, one has the correspondence
\be
\label{6}
 A_n \ra \hat P_n \equiv | n \rgl \lgl n | \;  .
\ee
Because of the projector orthogonality property
\be
\label{7}
  \hat P_m \hat P_n = \dlt_{mn} \hat P_n \; ,
\ee
the events with non-coinciding indices are also orthogonal,
\be
\label{8}
 A_m \bigcap A_n = \dlt_{mn} A_m \;  .
\ee
For a union of mutually orthogonal events, there is the correspondence
\be
\label{9}
 \bigcup_n A_n ~ \ra ~ \sum_n \hat P_n \;  ,
\ee
which results in the additivity of the probabilities:
\be
\label{10}
 p\left ( \bigcup_n A_n \right ) = \sum_n p(A_n) \;  .
\ee
The procedure described above is called the standard measurement [19].

\section{Operationally uncertain measurements}

Situations can exist when the result of a measurement is not well
defined, so that one cannot tell that a particular event has occurred, but it
is only known that some of the events $A_n$ could be realized. This is what
is called an uncertain, inconclusive, or generalized measurement [4,5].

Suppose the observed event $A$ is a set $\{A_n\}$ of possible events. Though
the events $A_m$ and $A_n$ are orthogonal for $m \neq n$, they form not a
standard union but an {\it uncertain union} that we shall denote as
\be
\label{11}
 A \equiv \{ A_n \} \equiv \biguplus_n A_n
\ee
in order to distinguish it from the standard union $\bigcup_n A_n$. To the
uncertain event $A$, there corresponds the wave function
\be
\label{12}
 | A \rgl = \sum_n a_n | n \rgl \;   ,
\ee
where $|a_n|^2$ play the role of weights for the events $A_n$. Instead of
the correspondence (9) for the standard union, we now have the correspondence
\be
\label{13}
 \biguplus_n A_n ~ \ra ~ \hat P_A \equiv | A \rgl \lgl A | \;  .
\ee

Generally, the proposition operator $\hat{P}_A$ is not a projector.
Since
$$
 \hat P_A \hat P_B = \lgl A | B \rgl \; | A \rgl \lgl B | \; ,
$$
it becomes an orthogonal projector only when functions (12) for different
$A$ and $B$ are normalized and orthogonal, which is not necessarily required.
This proposition operator is connected with the projectors $\hat{P}_n$ by
the relation
$$
 \hat P_A = \sum_n | a_n |^2 \hat P_n +
\sum_{m\neq n} a_m^* a_n | n \rgl \lgl m | \;  .
$$
The probability of the uncertain event $A$ reads as
\be
\label{14}
 p(A) = p\left ( \biguplus_n A_n \right ) = \sum_n | a_n |^2 p(A_n)
+ q(A) \;  ,
\ee
where the second term
\be
\label{15}
q(A) \equiv \sum_{m\neq n} a_m^* a_n \lgl m | \hat\rho | n \rgl
\ee
is caused by the interference of the uncertain subevents $A_n$ that are
called {\it modes}. As we see the probability of an uncertain union is not
additive, contrary to the probability of the standard union (10).

\section{Quantum joint probabilities}

To define the joint probability of different events, we follow the theory of
Ref. [16]. Let us be interested in two observables $\hat{A}$ and $\hat{B}$ that
can be commuting or not. Spanning the eigenfunctions of $\hat{A}$, we construct
a Hilbert space $\mathcal{H}_A$. Similarly, the observable $\hat{B}$ possesses
the eigenvalues and eigenfunctions given by the eigenproblem
\be
\label{16}
 \hat B | \al \rgl = B_\al | \al \rgl \;  .
\ee
Spanning the eigenfunctions of $\hat{B}$, we have a Hilbert space $\mathcal{H}_B$.
These two observables are treated as a tensor product $\hat{A} \bigotimes \hat{B}$
acting on the tensor-product Hilbert space
\be
\label{17}
 \cH_{AB} \equiv \cH_A \bigotimes \cH_B \;  .
\ee
The composite event of observing $A_n$ and $B_\alpha$ induces the correspondence
\be
\label{18}
 A_n \bigotimes B_\al ~ \ra ~ \hat P_n \bigotimes \hat P_\al \; ,
\ee
where
\be
\label{19}
 \hat P_\al \equiv | \al \rgl \lgl \al |
\ee
is a projector in $\mathcal{H}_B$.

The system state is now a statistical operator $\hat{\varrho}$ on the
tensor-product space (17). The joint probability of the composite event (18)
reads as
\be
\label{20}
  p(A_n \bigotimes B_\al ) = {\rm Tr} \hat \rho \hat P_n \bigotimes
\hat P_\al \equiv \lgl \hat P_n \bigotimes \hat P_\al \rgl \;  ,
\ee
with the trace over space (17). The composite event (18) is the simplest
composite event, which enjoys the factor form, being composed of two
elementary events. More complicated structures arise when at least one of
the events is a union. It is important to emphasize the difference between
the standard union and the uncertain union introduced in Sec. 3.

Considering the composite event, being a product of an elementary event $A_n$
and the standard union of mutually orthogonal events $B_\alpha$, we use the
known property [20] of composite events:
\be
\label{21}
  A_n \bigotimes \; \bigcup_\al B_\al =
\bigcup_\al A_n \bigotimes B_\al \; .
\ee
In the right-hand side of Eq. (21), we have the union of mutually orthogonal
composite events, since $B_\alpha$ are mutually orthogonal [20]. Hence
\be
\label{22}
 p \left ( A_n \bigotimes \bigcup_\al B_\al \right ) =
\sum_\al p(A_n \bigotimes B_\al ) \; .
\ee
Therefore the probability of a composite event, with one of the factors being
the standard union, is additive.

However, the situation is essentially different when dealing with an uncertain
union introduced in Sec. 3. Suppose we have such an uncertain union
\be
\label{23}
 B \equiv \{ B_\al \} \equiv \biguplus_\al B_\al
\ee
corresponding to a function
\be
\label{24}
 | B \rgl = \sum_\al b_\al | \al \rgl \; .
\ee
We may construct a composite event, or prospect
\be
\label{25}
 \pi_n =  A_n \bigotimes B = A_n \bigotimes \biguplus_\al B_\al \; ,
\ee
which corresponds to the prospect state
\be
\label{26}
  | \pi_n \rgl = | n \rgl \bigotimes | B \rgl =
\sum_{n\al} b_\al | n \al \rgl \; .
\ee
Then prospect (25) induces the correspondence
\be
\label{27}
\pi_n ~ \ra ~ \hat P(\pi_n) \equiv | \pi_n \rgl \lgl \pi_n |
\ee
defining the prospect operator $\hat{P}(\pi_n)$.

Note that the prospect states (26), in general, are not orthogonal and
normalized to one. Because of this, the prospect operators, generally, are
not projectors. However, the resolution of unity is required:
\be
\label{28}
 \sum_n \hat P(\pi_n) = \hat 1_{AB} \;  ,
\ee
where $\hat{1}_{AB}$ is the unity operator in space (17). The family
$\{\hat{P}(\pi_n)\}$ composes a positive operator-valued measure.

The prospect probability
$$
p(\pi_n) \equiv {\rm Tr} \hat\rho \hat P(\pi_n) \equiv
\lgl \hat P(\pi_n) \rgl \; ,
$$
with the trace over space (17), becomes the sum of two terms,
\be
\label{29}
 p(\pi_n) = f(\pi_n) + q(\pi_n) \;  ,
\ee
the first term
\be
\label{30}
 f(\pi_n) \equiv \sum_\al | b_\al |^2 p(A_n \bigotimes B_\al ) \; ,
\ee
containing diagonal elements with respect to $\alpha$, and the second term
\be
\label{31}
 q(\pi_n) \equiv \sum_{\al\neq\bt} b_\al^* b_\bt
\lgl n\al | \hat\rho | n \bt \rgl \;  ,
\ee
formed by the nondiagonal elements. By constructions and due to the resolution
of unity (28), the prospect probability (29) satisfies the properties
\be
\label{32}
 \sum_n p(\pi_n) = 1 \; , \qquad 0 \leq p(\pi_n) \leq 1 \;  ,
\ee
making the family $\{p(\pi_n)\}$ a probability measure.

Expression (31) is caused by the quantum nature of the considered events
producing interference of the modes composing the uncertain union (23).
Therefore quantity (31) can be called the {\it interference factor} or
{\it coherence factor}.

According to the quantum-classical correspondence principle, going back
to Bohr [21], classical theory is to be the limiting case of quantum theory,
when quantum effects vanish. In the present case, this implies that when
the interference (coherence) factor tends to zero, the quantum probability
has to tend to a classical probability. Keeping in mind this decoherence
process, we require the validity of the quantum-classical correspondence
principle in the form
\be
\label{33}
 p(\pi_n) ~ \ra ~ f(\pi_n) \; , \qquad q(\pi_n) ~ \ra ~ 0 \;  ,
\ee
assuming that the decoherence process leads to the classical probability
$f(\pi_n)$. Being a probability, this classical factor needs to be normalized,
so that to satisfy the conditions
\be
\label{34}
 \sum_n f(\pi_n) = 1 \; , \qquad 0 \leq f(\pi_n) \leq 1 \;  .
\ee
As a consequence of Eqs. (32) and (34), the interference factor enjoys the
properties
\be
\label{35}
 \sum_n q(\pi_n) = 0 \; , \qquad - 1 \leq q(\pi_n) \leq 1 \;  .
\ee

As is seen, the appearance of the interference factor is due to the structure
of the composite event (25) containing an uncertain union. The occurrence of
such a structure can be due to two reasons. One possibility can occur when
one accomplishes measurements with the observable $\hat{B}$, which provide
an uncertain result, and then realizes measurements with the observable
$\hat{A}$ leading to an operationally testable result. The second possibility
can correspond to an operationally testable measurement with $\hat{A}$, when
it is known that the measured system has been subject to some not well
controlled perturbations that could be of external or internal origin.
External perturbations can be caused by the influence of surrounding, and internal
perturbations can be produced by the measurer and measuring devices. In the
case when uncertainty is induced by uncontrolled perturbations, one may either
trust the results of the final measurement for $\hat{A}$, characterizing the
trust by $B_1$, or one may not trust these measurements, denoting this by $B_2$.
Then the uncertain union is $B_1 \biguplus B_2$. Since this uncertainty is not
caused by another measurement for another observable, but is due to uncontrolled
perturbations and the measurer trust, it can be termed {\it hidden uncertainty}.
Such a hidden uncertainty can also arise in quantum decision theory [11-15],
when a decision maker chooses between several lotteries whose setup is not
trustable, i.e.  dependent on uncertain factors.

\section{Average interference factor}

The interference factor is, of course, a contextual random quantity, depending
on the kind of measurements, measured observables, and experimental setup. But
we may try to evaluate its expected value.

Being a random quantity, the interference factor should be characterized by a
probability distribution $\varphi(q)$. The latter has to be normalized on the
interval $[-1,1]$, so that
\be
\label{36}
\int_{-1}^1 \vp(q)\; dq = 1 \;   .
\ee
And the alternation condition (35) can be written as
\be
\label{37}
 \int_{-1}^1 q \vp(q)\; dq = 0 \;  .
\ee
In the following, using the notation
\be
\label{38}
 q_+ \equiv \int_0^1 q \vp(q)\; dq \; , \qquad
q_- \equiv \int_{-1}^0 q \vp(q)\; dq \; ,
\ee
one has
\be
\label{39}
 q_+ + q_- = 0 \;  .
\ee
It is easy to see [12,15] that, for the case of non-informative prior, when
the distribution $\varphi(q)$ is uniform, we have the {\it quarter law},
when $q_+ = 1/4$, $q_- = -1/4$.

This result can be generalized showing that the quarter law is valid for
a wide class of distributions possessing some symmetry properties. As an
example, let us take the often used beta-distribution that we define here
on the interval $[-1,1]$ in the following way:
$$
 \vp(q) = \frac{\lbd_+}{B(\al,\bt)} \; q^{\al-1} ( 1 - q)^{\bt-1} \qquad
( 0 \leq q \leq 1 ) \;  ,
$$
\be
\label{40}
 \vp(q) = \frac{\lbd_-}{B(\mu,\nu)} \; |q |^{\mu-1} ( 1 - |q | )^{\nu-1}
\qquad ( -1 \leq q \leq 0 ) \;  ,
\ee
where the parameters $\alpha, \beta, \mu, \nu, \lambda_+, \lambda_-$ are
positive and
$$
 B(\al,\bt) = \frac{\Gm(\al)\Gm(\bt)}{\Gm(\al+\bt)} \; , \qquad
B(\mu,\nu) = \frac{\Gm(\mu)\Gm(\nu)}{\Gm(\mu+\nu)} \;   .
$$
The normalization condition (36) gives
$$
\lbd_+ + \lbd_ - = 1 \;  .
$$
The quantities (38) are given by
$$
 q_+ = \frac{\al\lbd_+}{\al+\bt} \; , \qquad
q_- = -\; \frac{\mu\lbd_-}{\mu+\nu} \;  .
$$
And equation (37) results in the equality
$$
 \frac{\al\lbd_+}{\al+\bt} =  \frac{\mu\lbd_-}{\mu+\nu} \; .
$$

Meanwhile, the values of $q_+$ and $q_-$ are not uniquely defined. But,
if we assume a symmetric distribution, such that $\lambda_+ = \lambda_- = 1/2$
and $\alpha = \beta$, then it follows that $\mu = \nu$ and the quarter law
is valid for arbitrary positive $\alpha$ and $\mu$:
$$
 q_+ = \frac{1}{4}\; , \qquad  q_- = -\; \frac{1}{4}\; .
$$
The non-informative prior is a particular case of expression (\ref{40}),
when $\alpha = \beta = 1, \mu = \nu = 1$, which yields
$\lambda_+ = \lambda_- = 1/2$ and gives the uniform distribution
$\varphi(q) = 1/2$.

\section{Prospect and state entanglement}

Entanglement plays an important role in quantum information processing [1]
and has thus been widely studied for many physical systems of different
nature (see, e.g., Refs. [22-27]).

As is demonstrated in Sec. 3, the interference factor arises if the considered
prospect is composite and contains an uncertain union, as in prospect (25).
This prospect can be termed entangled. It is principally different from the
composite prospect (21) that yields the additive probability (22) containing
no interference. Thus the existence of an entangled prospect is a necessary
condition for the appearance of an interference factor.

Another necessary condition is that the system state $\hat{\varrho}$ should also be
entangled. To show that a disentangled state does not produce interference,
let us take the system state in the disentangled product form
\be
\label{41}
 \hat \rho = \hat \rho_A \bigotimes \hat \rho_B \;  .
\ee
Then the interference factor (31) becomes
\be
\label{42}
q(\pi_n) = \sum_{\al\neq\bt} b_\al^* b_\bt \lgl n | \hat\rho_A | n \rgl
\lgl \al | \hat \rho_B | \bt \rgl \;   .
\ee
In view of property (35) and taking into account the normalization
\be
\label{43}
{\rm Tr}_A \hat\rho_A = \sum_n  \lgl n | \hat\rho_A | n \rgl = 1 \;  ,
\ee
we get
\be
\label{44}
 \sum_n q(\pi_n) =
\sum_{\al\neq\bt} b_\al^* b_\bt \lgl \al | \hat\rho_B | \bt \rgl = 0 \; .
\ee
Using Eqs. (42) and (44), we find
\be
\label{45}
 q(\pi_n) = \lgl n | \hat\rho_A | n \rgl \sum_n q(\pi_n) = 0 \;  .
\ee
So, the disentangled state (41) does not allow for a nontrivial interference
factor.

Though the state entanglement is a necessary condition for the occurrence of
interference, it is not sufficient. As a counterexample, we can consider a
maximally entangled state such as the multimode state
\be
\label{46}
\hat\rho \equiv | \psi \rgl \lgl \psi | =
\frac{1}{M} \sum_{mn} | mm \rgl \lgl nn |
\ee
composed of the multimode function
\be
\label{47}
 | \psi \rgl = \frac{1}{\sqrt{M} } \; \sum_m | mm \rgl \; ,
\ee
with the number of modes $M$,
\be
\label{48}
M = \sum_m  1 = {\rm dim}\cH_A = {\rm dim} \cH_B \;  .
\ee
The entanglement-production measure [28,29] for the statistical operator (46) is
\be
\label{49}
 \ep(\hat\rho ) = \log M \;  .
\ee
But the interference factor (31) is zero:
\be
\label{50}
 q(\pi_n) = \frac{1}{M}
\sum_{\al\neq\bt} b_\al^* b_\bt \dlt_{\al n} \dlt_{\bt n} = 0 \;  .
\ee

In this way, the entanglement of the prospect and of the system state is a
necessary, but not sufficient, condition for the occurrence of the mode
interference.

\section{Two-mode Bose system}

As an illustration of the approach, let us consider a gas of trapped Bose
atoms at ultracold temperatures, when practically all atoms pile down to
a Bose-Einstein condensed state. At the temperatures close to zero and
in the presence of very weak interactions, the trapped Bose system can be described by the
nonlinear Schr\"{o}dinger (NLS) equation with a discrete spectrum due to
the atomic confinement [30,31]. The functions that are the solutions to
the stationary NLS equation are termed the coherent modes. In equilibrium,
all atoms settle down to the collective ground-state energy level. By
applying an alternating external field, which either modulates the trapping
potential or induces the oscillation of the scattering length by means of
Feshbach resonance techniques [32], the atomic gas goes into a nonequilibrium state.
Tuning the frequency of the modulating field to the resonance with a
transition frequency between the atomic energy levels, it is feasible to
generate one or several excited modes. Thus it is possible to create a
two-mode Bose system [30,31]. Other ways of generating two-mode (or multimode)
Bose systems could be by splitting condensates with an external beam [33]
or loading condensates into a double-well (or multi-well) potential,
where the ground-state energy level splits into two (or several) energy
levels, with the splitting being regulated by adjusting the parameters
of the potential [34].

Let us denote the modes, that are the normalized solutions to the stationary
NLS equation, as $|n \rangle$. An event $A_n$ symbolizes the observation of
a $|n \rangle$-th mode. The probabilities of separate events $A_n$ are
calculated in the standard way, as in Sec. 2. But our aim is to show how one
could define the joint probabilities in the case of entangled prospects of
type (25). The prospects of this type can arise owing to two reasons.

One possibility corresponds to the situation when, at a given time, one
accomplishes a measurement for observing events $A_n$ while, at the previous
times, realizing uncertain measurements for the events denoted as $B_\alpha$.
The measurements can be treated as nondestructive, yielding only phase shifts,
but not directly influencing the level populations [35].

The other interpretation could be as follows. At a given time, one accomplishes
measurements for observing events $A_n$, while one is aware that, at the
previous times, the system has been subject to uncontrolled perturbations.
It is clear that these two cases are analogous to each other, since uncertain
measurements can be mathematically described as uncontrolled perturbations.

According to Sec. 4, the interference factor, caused by the interference
of modes in the process of uncertain measurements, is
\be
\label{51}
 q(\pi_n) = p(\pi_n) - f(\pi_n) \;  .
\ee
Resorting to the channel-state duality, the picture can be translated into
a temporal representation [16,35]. Then, for the prospect probability, we have
the correspondence
\be
\label{52}
p(\pi_n) ~ \ra ~ p_n(t) \;  ,
\ee
respectively, for the interference factor,
\be
\label{53}
  q(\pi_n) ~ \ra ~ q_n(t) \; ,
\ee
where the quantities $p_n(t)$ and $q_n(t)$ are defined through a channel
picture. The influence of measurements can be described as the action of a
random noise of strength $\sigma$. Without the noise, entangling modes, the
probability is given by
\be
\label{54}
 f_n(t) \equiv \lim_{\sgm\ra 0} p_n(t) \;  .
\ee
Then, the interference factor can be defined as
\be
\label{55}
 q_n(t) = p_n(t) - f_n(t) \;  .
\ee

Considering a Bose-condensed system with two modes, generated by means of
the trap modulation [30,31], and keeping in mind nondestructive measurements
[35], we have the mode populations
\be
\label{56}
 p_1(t) = \frac{1-s(t)}{2} \; , \qquad p_2(t) = \frac{1+s(t)}{2} \;   ,
\ee
in which $s(t)$ is the population imbalance, satisfying the equation of motion
\be
\label{57}
  \frac{ds}{dt} = - b \; \sqrt{1 - s^2} \; \sin x \; ,
\ee
and $x(t)$ is the phase difference described by the equation
\be
\label{58}
 dx = s \left ( 1 + \frac{b}{\sqrt{1-s^2}} \; \cos x \right ) dt
+ \sgm d W_t \; .
\ee
Here $b$ is a parameter quantifying the amplitude of the pumping field
modulating the trap, $W_t$ is the standard Wiener process with the standard
deviation $\sigma$, and time $t$ is dimensionless.

In the absence of random perturbations ($\sgm =0$), there exist two regimes in the
dynamics of Eqs. (57) and (58), depending on the value of the pumping
amplitude $b$ with respect to the critical value
\be
\label{59}
 b_c = \frac{s_0^2}{2(1+\sqrt{1-s_0^2}\; \cos x_0 ) } \;  ,
\ee
where $s_0 \equiv s(0), x_0 \equiv x(0)$ are the initial conditions. The
subcritical regime, when $b < b_c$, corresponds to the Rabi oscillations
of $s(t)$, while the supercritical regime, when $b > b_c$, is the regime
of Josephson oscillations [30,31]. The value of $b_c$ varies in the
diapason
\be
\label{60}
0 \leq b_c \leq \frac{1}{2} \qquad ( 0 \leq s_0^2 \leq 1 ) \; .
\ee

\begin{figure}[ht]
\begin{minipage}{16pc}
\includegraphics[width=16pc]{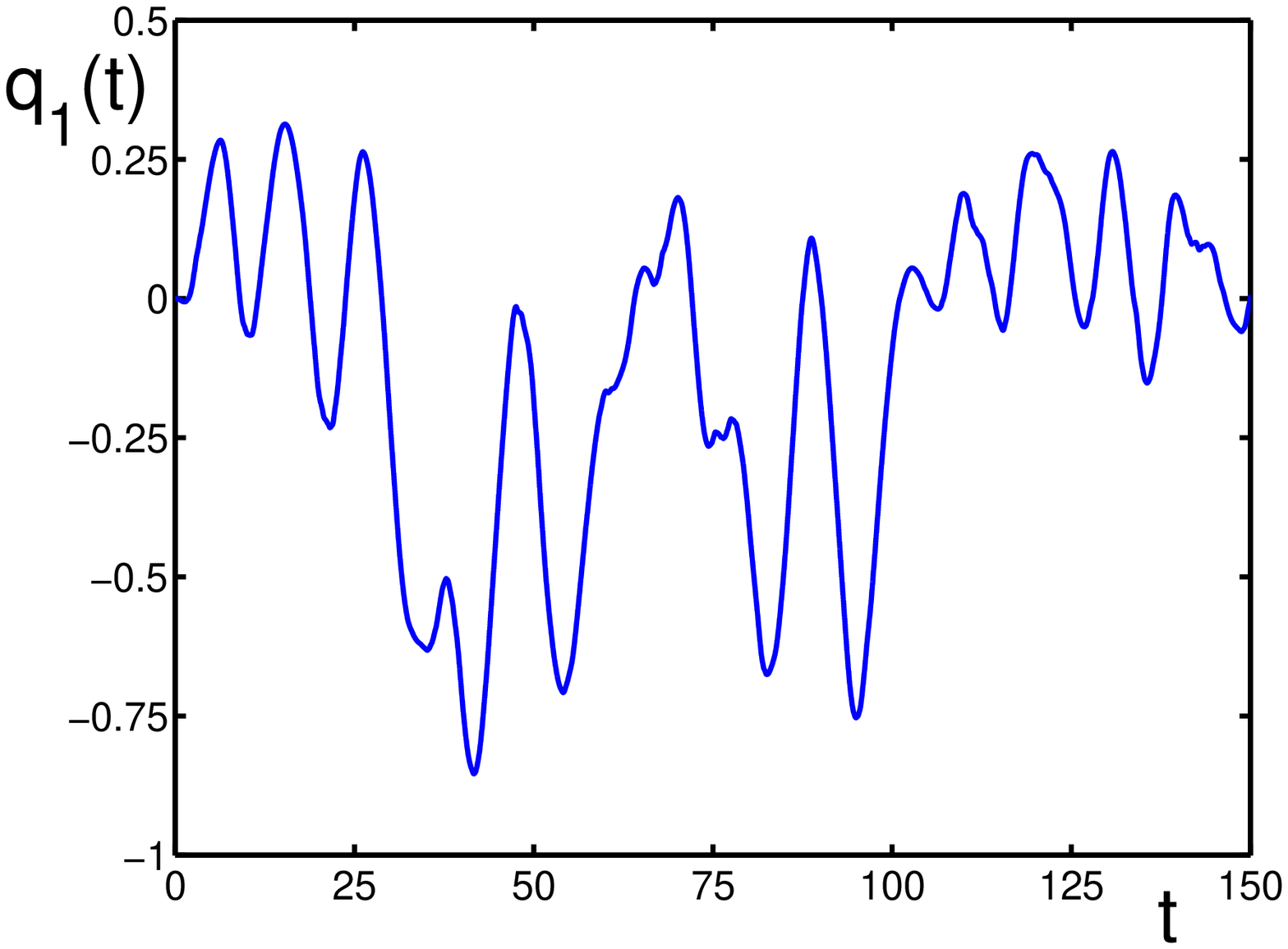}
\caption{\label{l1}Interference factor $q_1(t)$ as a function of dimensionless
time for the subcritical regime, with $b = 0.25 < b_c = 0.282$.}
\end{minipage}\hspace{5pc}%
\begin{minipage}{16pc}
\includegraphics[width=16pc]{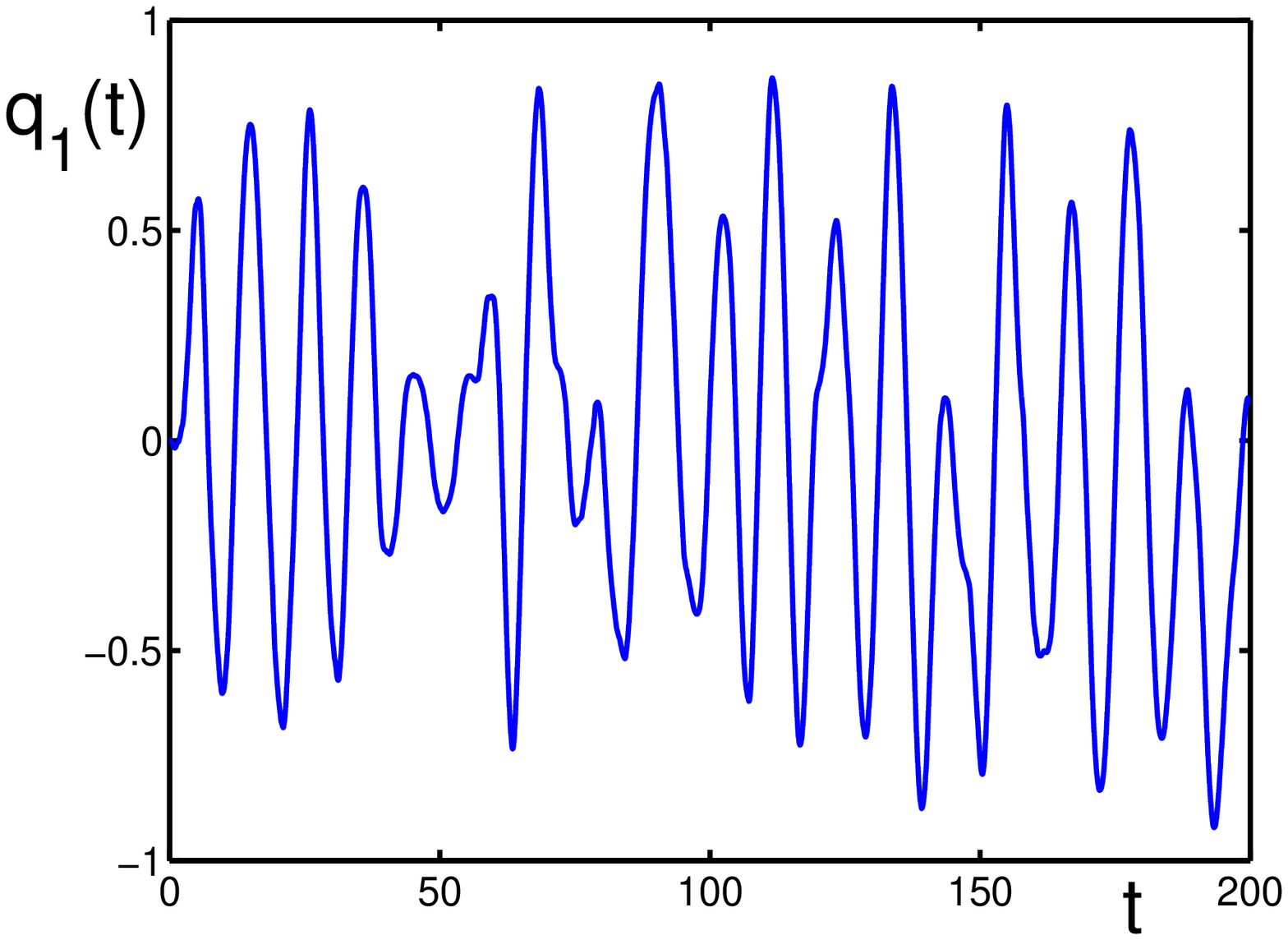}
\caption{\label{l2}Interference factor $q_1(t)$ as a function of dimensionless
time for the supercritical regime, with $b = 0.5 > b_c = 0.282$.}
\end{minipage} 
\end{figure}

Solving Eqs. (57) and (58) in the presence of the random perturbations,
we take the initial conditions $s_0 = -0.9, x_0 = 0$, which define the
critical pumping parameter $b_c = 0.282$. The results of the numerical
solution for the interference factor (55), with $n=1$, are presented in
Fig. 1 for the subcritical regime and in Fig. 2 for supercritical regime.
It is sufficient to present only one interference factor, since the second
is given by $q_2(t) = - q_1(t)$. 

One can observe that the fluctuations of the interference
factor are larger in the supercritical regime.

In conclusion, we have shown how the quantum probabilities for composite
events, related to quantum measurements, can be defined. A special attention
has been payed to the case of operationally uncertain measurements, when there
appears an interference factor. We described how the average value of the
interference factor can be estimated under rather general conditions. The
necessity of entanglement in the considered prospect, as well as in the system
state, for the appearance of interference has been stressed. An example of calculating
the interference factor, employing the channel-state duality, for a trapped
Bose-Einstein condensate with generated coherent modes, was demonstrated.

\ack
Financial supports from the Swiss National Foundation and from the
Russian Foundation for Basic Research are acknowledged.

\end{document}